\documentclass[12pt]{article}

\catcode`@=12
\begin{document}
\thispagestyle{empty}
\begin{flushright} 
UCRHEP-T375\\ 
IPPP-04-20\\
DCPT-04-40\\
May 2004\
\end{flushright}
\vspace{1.0in}
\begin{center}
{\LARGE	\bf Non-Abelian Discrete Symmetries\\ and Neutrino Masses: 
Two Examples\\}
\vspace{1.5in}
{\bf Ernest Ma\\}
\vspace{0.2in}
{\sl Physics Department, University of California, Riverside, 
California 92521, USA and\\ Institute for Particle Physics Phenomenology,
Physics Department, University of Durham, Durham DH1 3LE, UK}
\vspace{1.5in}
\end{center}
\begin{abstract}\
Two recent examples of non-Abelian discrete symmetries ($S_3$ and $A_4$) 
in understanding neutrino masses and mixing are discussed.
\end{abstract}

\newpage
\baselineskip 24pt

\section{Introduction}

In the standard model of quark and lepton interactions, quark and 
charged-lepton masses come from the Yukawa couplings of the left-handed 
doublets $(u,d)_L$ and $(\nu,l)_L$ with the right-handed singlets $u_R$, 
$d_R$, and $l_R$ through the vacuum expectation value of the one scalar 
Higgs doublet $(\phi^+,\phi^0)$.  The quark mixing matrix $V_{CKM}$ is 
then obtained from the mismatch in the diagonalization of the $up$ and 
$down$ quark mass matrices.  Remarkably, $V_{CKM}$ turns out to be 
almost identically the unit matrix, i.e. quark mixing angles are all small. 
[For 3 families, $V_{CKM}$ has 3 angles and 1 phase.]  On the other hand, 
the analogous $U_{MNS(P)}$ which is obtained from the mismatch in the 
diagonalization of the charged-lepton mass matrix and that of the neutrino 
mass matrix, is far from being the unit matrix. Whereas one angle is indeed 
small, the other two are definitely large.  Indeed, to a good first 
approximation, 
\begin{equation}
U_{MNS(P)} \simeq \pmatrix {\sqrt{2/3} & 1/\sqrt 3 & 0 \cr -1/\sqrt 6 & 
1/\sqrt 3 & -1/\sqrt 2 \cr -1/\sqrt 6 & 1/\sqrt 3 & 1/\sqrt 2}.
\end{equation}
In the convention
\begin{equation}
U_{MNS(P)} = \pmatrix {1 & 0 & 0 \cr 0 & c_{23} & -s_{23} \cr 0 & s_{23} & 
c_{23}} \pmatrix {c_{13} & 0 & s_{13} e^{-i\delta} \cr 0 & 1 & 0 \cr -s_{13} 
e^{i\delta} & 0 & c_{13}} \pmatrix {c_{12} & -s_{12} & 0 \cr s_{12} & c_{12} 
& 0 \cr 0 & 0 & 1},
\end{equation}
this means that
\begin{equation}
\theta_{23} \simeq \pi/4, ~~~ \theta_{12} \simeq \tan^{-1} (-1/\sqrt2), ~~~ 
\theta_{13} \simeq 0, 
\end{equation}
which are consistent with the present experimental constraints \cite{venice}:
\begin{equation}
\sin^2 2 \theta_{23} > 0.91~(90\%~{\rm C.L.}), ~~ 0.30 < \tan^2 \theta_{12} < 
0.52~(90\%~{\rm C.L.}), ~~ \sin^2 \theta_{13} < 0.067~(3\sigma).
\end{equation}

There are a number of approaches in trying to understand the origin of 
quark and lepton mass matrices.  Most attempts want to relate mixing angles 
with mass ratios.  This was historically motivated by the phenomenologically 
successful ansatz $\theta_C \simeq \sqrt {m_d/m_s}$ for the Cabibbo angle. 
One often assumes that there is a symmetry behind this relationship. That 
may be so, but a better question to ask is perhaps whether or not there 
exists a family symmetry which tells us that $V_{CKM}=1$ and $U_{MNS(P)} 
\neq 1$. Obviously, if each family has its own Abelian (continuous or 
discrete) symmetry, then there is no mixing among families. That works 
well for $V_{CKM}$ but not $U_{MNS(P)}$. If neutrino masses are purely 
Dirac, then the analogous structure of the quark and lepton sectors would 
definitely rule out the existence of such a family symmetry. However, 
if neutrino masses are Majorana, then it is indeed possible to have 
$V_{CKM} = 1$ and $U_{MNS(P)}$ as given by Eq.~(1) as the result of a 
symmetry, as shown below. To fit experimental data, small (radiative) 
corrections are needed from physics beyond the Standard Model.

\section{$S_3$ for Two Families}

\subsection{Representations of $S_3$}

The group of permutations of 3 objects is $S_3$.  It is isomorphic to the 
group of three-dimensional rotations of an equilateral triangle to itself, 
i.e. the dihedral group $D_3$. It has 6 elements and 3 irreducible 
representations: \underline {1}, \underline {1}$'$, and \underline {2}. 
As such, it is ideal for describing 2 families.

Since \underline {1}$' \times$ \underline {1}$'$ = \underline {1}, it is 
clear that a field transforming as \underline {1}$'$ should have a $Z_2$ 
parity of $-1$, i.e. $\phi \to -\phi$.  On the other hand, a doublet 
$(\phi_1,\phi_2)$ under $S_3$ has a choice of representations as long as 
it satisfies
\begin{equation}
\underline {2} \times \underline {2} = \underline {1} + \underline {1}' 
+ \underline {2}.
\end{equation}
Different representations are simply related by a unitary tranformation. 
The most convenient representation of $S_3$ is a complex representation 
\cite{ma91,desh92} such that the products of the doublets $\phi_{1,2}$ 
and $\psi_{1,2}$ are given by
\begin{eqnarray}
&& \phi_1 \psi_2 + \phi_2 \psi_1 \sim \underline {1}, \\ 
&& \phi_1 \psi_2 - \phi_2 \psi_1 \sim \underline {1}', \\
&& (\phi_2 \psi_2 , \phi_1 \psi_1) \sim \underline {2}.
\end{eqnarray}
Note that $\phi_1 \phi_1^* + \phi_2 \phi_2^*$ is an invariant.  Hence 
$(\phi_2^*,\phi_1^*)$ is a doublet.  Note also that $S_3$ has the special 
property that the symmetric product of 3 doublets, i.e. $\phi_1 \psi_1 \chi_1 
+ \phi_2 \psi_2 \chi_2$ is a singlet.

Specifically, the 6 group elements are the identity: $e$, the cyclic and 
anti-cyclic permutations of three objects: $g_c$ and $g_a$, and the three 
interchanges of two objects leaving the third fixed: $g_1$, $g_2$, $g_3$. 
Their representation matrices are [1,1,1,1,1,1] in \underline {1}, 
$[1,1,1,-1,-1,-1]$ in \underline {1}$'$, and
\begin{equation}
\left[ \pmatrix {1 & 0 \cr 0 & 1}, ~~\pmatrix {\omega & 0 \cr 0 & \omega^2}, 
~~\pmatrix {\omega^2 & 0 \cr 0 & \omega}, ~~\pmatrix {0 & \omega^2 \cr 
\omega & 0}, ~~\pmatrix {0 & \omega \cr \omega^2 & 0}, ~~\pmatrix {0 & 1 \cr 
1 & 0} \right]
\end{equation}
in \underline {2} respectively for [$e$, $g_c$, $g_a$, $g_1$, $g_2$, $g_3$], 
where $\omega = e^{2\pi i/3}$.

\subsection{Quarks and Leptons under $S_3$}

Consider a world of only two (i.e. the second and third) families of quarks 
and leptons.  Choosing the convention that all fermions are left-handed, a 
natural assignment is 
\cite{cfm04}
\begin{eqnarray}
&& Q_i = (u_i,d_i), ~L_i = (\nu_i, l_i) \sim \underline {2}, ~(i=2,3) \\ 
&& u^c_2, ~d^c_2, ~l^c_2 \sim \underline {1}, ~~~u^c_3, ~d^c_3, ~l^c_3 
\sim \underline {1}'.
\end{eqnarray}
To allow $u$, $d$, and $l$ to have Dirac mass terms, two scalar electroweak 
doublets $(\phi_i^0, \phi_i^-) ~(i=2,3)$ transforming as an $S_3$ doublet are 
required.  The invariant leptonic Yukawa couplings are then
\begin{equation}
{\cal L}_Y = f_2 (\phi_2 L_3 + \phi_3 L_2) l^c_2 + f_3 (\phi_2 L_3 - \phi_3 
L_2) l^c_3 + H.c.,
\end{equation}
resulting in the $2 \times 2$ mass matrix linking $l_{2,3}$ to $l^c_{2,3}$ 
below:
\begin{equation}
{\cal M}_{ll^c} = \pmatrix {f_2 v_3 & -f_3 v_3 \cr f_2 v_2 & f_3 v_2},
\end{equation}
where $v_i = \langle \phi^0_i \rangle$. 
On the other hand, the Majorana neutrino mass matrix depends on the 
product of $L_i$ and $L_j$, i.e. Eq.~(5).  Thus a choice of scalar 
representations is available.  Suppose two scalar triplets $(\xi_i^{++}, 
\xi_i^+,\xi_i^0) ~(i=2,3)$ transforming as an $S_3$ doublet are used, then
\begin{equation}
{\cal L}_Y = h (L_2 L_2 \xi_2 + L_3 L_3 \xi_3) + H.c.,
\end{equation}
which leads to
\begin{equation}
{\cal M}_\nu = \pmatrix {h u_2 & 0 \cr 0 & h u_3},
\end{equation}
where $u_i = \langle \xi^0_i \rangle$. Comparing Eqs.~(13) and (15), the 
lepton mixing matrix is easily obtained for $v_2=v_3=v$, because Eq.~(13) is 
diagonalized by
\begin{equation}
U^\dagger = {1 \over \sqrt 2} \pmatrix {1 & 1 \cr -1 & 1}
\end{equation}
on the left and the unit matrix on the right, which of course means maximal 
mixing. The charged-lepton mass eigenvalues are then $\sqrt 2 f_2 v$ and 
$\sqrt 2 f_3 v$, whereas the Majorana neutrino mass eigenvalues are $h u_2$ 
and $h u_3$.  They can all be different and yet maximal mixing is ensured. 
This depends of course on the condition $v_2=v_3$ which can be maintained 
in the Higgs potential by the interchange symmetry $\phi_2 \leftrightarrow 
\phi_3$. The trilinear scalar couplings $\mu_2 \phi_2 \phi_3 \xi_2$ and 
$\mu_3 \phi_2 \phi_3 \xi_3$ break $S_3$ softly but are invariant under 
$\phi_2 \leftrightarrow \phi_3$.  Hence $\mu_2 \neq \mu_3$ would imply 
$u_2 \neq u_3$. Note that since $m_\tau >> m_\mu$ means $f_3 >> f_2$, 
the charged-lepton matrix is diagonalized by
\begin{equation}
U^\dagger = {1 \over \sqrt{v_2^2+v_3^2}} \pmatrix {v_2 & v_3 \cr -v_3 & v_2}
\end{equation}
to a good approximation even if $v_2 \neq v_3$. In that case, the mixing 
angle is given by $\tan^{-1} (v_3/v_2)$, which can differ from $\pi/4$.

In the quark sector, the $down$ quark mass matrix has the same form as 
Eq.~(13), i.e.
\begin{equation}
{\cal M}_{dd^c} = \pmatrix {f^d_2 v_3 & -f^d_3 v_3 \cr f^d_2 v_2 & f^d_3 v_2},
\end{equation}
but the $up$ quark mass matrix is given by
\begin{equation}
{\cal M}_{uu^c} = \pmatrix {f^u_2 v_2^* & -f^u_3 v_2^* \cr f^u_2 v_3^* & 
f^u_3 v_3^*}
\end{equation}
instead, because $(\phi_3^*,\phi_2^*)$ must be used in place of 
$(\phi_2,\phi_3)$.  This means that there is now a mismatch between the 
two diagonalized mass matrices and the mixing angle is given by \cite{cfm04}
\begin{equation}
\theta_q = 2 \left[ {\pi \over 4} - \tan^{-1} (v_3/v_2) \right] = 
{\pi \over 2} - 2 \theta_l.
\end{equation}
In this way, the smallness of the quark mixing between the second and third 
families is related to the deviation from maximal mixing in the $\mu-\tau$ 
sector. To include the first family of quarks and leptons, $S_3$ singlets 
must be used.  Since either \underline {1} or \underline {1}$'$ must be 
chosen, there has to be mixing of the first family into the $2-3$ sector, 
but its exact form or magnitude cannot be fixed by $S_3$ alone. For a 
specific successful application, see Ref.~[4].

\section{$A_4$ for Three Families}

\subsection{Representations of $A_4$}

The group of even permutations of 4 objects is $A_4$.  It is isomorphic to 
the group of three-dimensional rotations of a regular tetrahedron, one of 
five perfect geometric solids known to the ancient Greeks and identified 
by Plato with the element ``fire''.  It is thus a discrete subgroup of 
$SO(3)$.  It is also isomorphic to $\Delta(12)$ which is a discrete subgroup 
of SU(3) \cite{disc_sg}.  It has 12 elements and 4 irreducible reprsentations: 
\underline {1}, \underline {1}$'$, \underline {1}$''$, and \underline {3}, 
with the multiplication rule
\begin{equation}
\underline {3} \times \underline {3} = \underline {1} + \underline {1}' + 
\underline {1}'' + \underline {3} + \underline {3},
\end{equation}
in analogy to Eq.~(5) for $S_3$.  As such, it is ideal for describing three 
families.  Specifically, the products of the triplets 
$\phi_{1,2,3}$ and $\psi_{1,2,3}$ are given by \cite{mr01}
\begin{eqnarray}
&& \phi_1 \psi_1 + \phi_2 \psi_2 + \phi_3 \psi_3 \sim \underline {1}, \\ 
&& \phi_1 \psi_1 + \omega^2 \phi_2 \psi_2 + \omega \phi_3 \psi_3 \sim 
\underline {1}', \\ 
&& \phi_1 \psi_1 + \omega \phi_2 \psi_2 + \omega^2 \phi_3 \psi_3 \sim 
\underline {1}'', \\ 
&& (\phi_2 \psi_3 , \phi_3 \psi_1, \phi_1 \psi_2) \sim \underline {3}, \\
&& (\phi_3 \psi_2 , \phi_1 \psi_3, \phi_2 \psi_1) \sim \underline {3},
\end{eqnarray}
where $\omega = e^{2 \pi i/3}$.  Note that $A_4$ also has the special property 
that the symmetric product of 3 triplets, i.e.
\begin{equation}
\phi_1 \psi_2 \chi_3 + \phi_1 \psi_3 \chi_2 + \phi_2 \psi_1 \chi_3 + \phi_2 
\psi_3 \chi_1 + \phi_3 \psi_1 \chi_2 + \phi_3 \psi_2 \chi_1
\end{equation}
is a singlet. 

Specifically, the 12 group elements are divided into 4 equivalence classes: 
$C_1$ contains only the identity, $C_2$ has 4 elements of order 3, $C_3$ 
also has 4 elements of order 3, and $C_4$ has 3 elements of order 2. 
The representation matrices in \underline {3} are given by
\begin{eqnarray}
C_1 &:& \pmatrix {1 & 0 & 0 \cr 0 & 1 & 0 \cr 0 & 0 & 1}, \\ 
C_2 &:& \pmatrix {0 & 0 & 1 \cr 1 & 0 & 0 \cr 0 & 1 & 0}, 
\pmatrix {0 & 0 & 1 \cr -1 & 0 & 0 \cr 0 & -1 & 0}, 
\pmatrix {0 & 0 & -1 \cr 1 & 0 & 0 \cr 0 & -1 & 0}, 
\pmatrix {0 & 0 & -1 \cr -1 & 0 & 0 \cr 0 & 1 & 0}, \\ 
C_3 &:& \pmatrix {0 & 1 & 0 \cr 0 & 0 & 1 \cr 1 & 0 & 0}, 
\pmatrix {0 & 1 & 0 \cr 0 & 0 & -1 \cr -1 & 0 & 0}, 
\pmatrix {0 & -1 & 0 \cr 0 & 0 & 1 \cr -1 & 0 & 0}, 
\pmatrix {0 & -1 & 0 \cr 0 & 0 & -1 \cr 1 & 0 & 0}, \\ 
C_4 &:& \pmatrix {1 & 0 & 0 \cr 0 & -1 & 0 \cr 0 & 0 & -1}, 
\pmatrix {-1 & 0 & 0 \cr 0 & 1 & 0 \cr 0 & 0 & -1}, 
\pmatrix {-1 & 0 & 0 \cr 0 & -1 & 0 \cr 0 & 0 & 1}.
\end{eqnarray}
They are [1,1,1,1], $[1,\omega,\omega^2,1]$, and 
$[1,\omega^2,\omega,1]$ in \underline{1}, \underline {1}$'$, and 
\underline {1}$''$ respectively.

\subsection{Quarks and Leptons under $A_4$}

In analogy to Eqs.~(10) and (11) for $S_3$, a natural assignment for 3 
families of quarks and leptons is \cite{mr01,ma02,bmv03,hrsvv,ma04}
\begin{eqnarray}
&& Q_i = (u_i,d_i), ~L_i = (\nu_i, l_i) \sim \underline {3}, ~(i=1,2,3) \\ 
&& u^c_1, ~d^c_1, ~l^c_1 \sim \underline {1}, ~~~u^c_2, ~d^c_2, ~l^c_2 
\sim \underline {1}', ~~~u^c_3, ~d^c_3, ~l^c_3 \sim \underline {1}''.
\end{eqnarray}
To allow $u$, $d$, and $l$ to have Dirac mass terms, three scalar electroweak 
doublets $(\phi_i^0, \phi_i^-) ~(i=1,2,3)$ transforming as an $A_4$ triplet 
are required.  The invariant leptonic Yukawa couplings are then
\begin{eqnarray}
{\cal L}_Y &=& f_1 (L_1 \phi_1 + L_2 \phi_2 + L_3 \phi_3) l_1^c \nonumber \\ 
&+& f_2 (L_1 \phi_1 + \omega L_2 \phi_2 + \omega^2 L_3 \phi_3) l_2^c 
\nonumber \\ 
&+& f_3 (L_1 \phi_1 + \omega^2 L_2 \phi_2 + \omega L_3 \phi_3) l_3^c + H.c.,
\end{eqnarray}
where \underline {1}$' \times$ \underline {1}$''$ = \underline {1} has been 
used.  The resulting $3 \times 3$ mass matrix linking $l_{1,2,3}$ to 
$l^c_{1,2,3}$ is
\begin{equation}
{\cal M}_{ll^c} = \pmatrix {f_1 v_1 & f_2 v_1 & f_3 v_1 \cr f_1 v_2 & f_2 
\omega v_2 & f_3 \omega^2 v_2 \cr f_1 v_3 & f_2 \omega^2 v_3 & f_3 \omega v_3},
\end{equation}
which is diagonalized simply by
\begin{equation}
U_L^\dagger = {1 \over \sqrt 3} \pmatrix {1 & 1 & 1 \cr 1 & \omega & \omega^2 
\cr 1 & \omega^2 & \omega}
\end{equation}
on the left and the unit matrix on the right for $v_1 = v_2 = v_3 = v$. 
The charged-lepton mass eigenvalues are then $\sqrt 3 f_1 v$, $\sqrt 3 f_2 v$, 
and $\sqrt 3 f_3 v$, which are of course free to be chosen as $m_e$, $m_\mu$, 
and $m_\tau$.  Since this matrix also diagonalizes the $up$ and $down$ 
quark mass matrices, the resulting quark mixing matrix is just the unit 
matrix, i.e. $V_{CKM} = 1$.

\subsection{Three Degenerate Neutrino Masses}

Consider now the $3 \times 3$ Majorana neutrino mass matrix.  Since the 
product of $L_i$ and $L_j$ is given by Eq.~(21), a choice of scalar 
representations is available, as in the case of $S_3$ discussed earlier. 
The simplest choice is to have one scalar triplet $(\xi_1^{++},\xi_1^+,
\xi_1^0)$ transforming as \underline {1} under $A_4$.  In that case,
\begin{equation}
{\cal L}_Y = h_1 (L_1 L_1 + L_2 L_2 + L_3 L_3) \xi_1 + H.c.,
\end{equation}
resulting in three degenerate neutrino masses, i.e.
\begin{equation}
{\cal M}_\nu = \pmatrix {m_0 & 0 & 0 \cr 0 & m_0 & 0 \cr 0 & 0 & m_0},
\end{equation}
where $m_0 = 2 h_1 \langle \xi_1^0 \rangle$.  In the $(e,\mu,\tau)$ basis, 
it becomes
\begin{equation}
{\cal M}_\nu^{(e,\mu,\tau)} = U_L^\dagger {\cal M}_\nu U_L^* = 
\pmatrix {m_0 & 0 & 0 \cr 0 & 0 & m_0 \cr 0 & m_0 & 0}.
\end{equation}

From the high scale where $A_4$ is broken to the electroweak scale, 
one-loop radiative corrections will change Eq.~(39) to
\begin{equation}
\pmatrix {m_0 & 0 & 0 \cr 0 & 0 & m_0 \cr 0 & m_0 & 0} +  R \pmatrix {m_0 
& 0 & 0 \cr 0 & 0 & m_0 \cr 0 & m_0 & 0} + \pmatrix {m_0 & 0 & 0 \cr 0 & 0 
& m_0 \cr 0 & m_0 & 0} R^T,
\end{equation}
where the radiative correction matrix is assumed to be of the most general 
form, i.e.
\begin{equation}
R = \pmatrix {r_{ee} & r_{e\mu} & r_{e\tau} \cr r_{e\mu}^* & r_{\mu\mu} & 
r_{\mu\tau} \cr r_{e\tau}^* & r_{\mu\tau}^* & r_{\tau\tau}}.
\end{equation}
Thus the observed neutrino mass matrix is given by
\begin{equation}
{\cal M}_\nu = m_0 \pmatrix {1+2r_{ee} & r_{e\tau} + r_{e\mu}^* & r_{e\mu} + 
r_{e\tau}^* \cr r_{e\mu}^* + r_{e\tau} & 2r_{\mu\tau} & 1+r_{\mu\mu}+
r_{\tau\tau} \cr r_{e\tau}^* + r_{e\mu} & 1+r_{\mu\mu}+r_{\tau\tau} & 
2r_{\mu\tau}^*}.
\end{equation}
Then using the redefinitions:
\begin{eqnarray}
&& \delta_0 \equiv r_{\mu\mu} + r_{\tau\tau} - r_{\mu\tau} - r^*_{\mu \tau}, 
\\ 
&& \delta \equiv 2r_{\mu\tau}, \\
&& \delta' \equiv r_{ee} - {1 \over 2} r_{\mu\mu} - {1 \over 2} r_{\tau\tau} 
- {1 \over 2} r_{\mu\tau} - {1 \over 2} r^*_{\mu \tau}, \\
&& \delta'' \equiv r_{e\mu}^* + r_{e\tau},
\end{eqnarray}
it becomes
\begin{equation}
{\cal M}_\nu = m_0 \pmatrix{1+\delta_0+\delta+\delta^*+2\delta' & \delta'' & 
\delta''^* \cr \delta'' & \delta & 1+\delta_0+(\delta+\delta^*)/2 \cr 
\delta''^* & 1+\delta_0+(\delta+\delta^*)/2 & \delta^*}.
\end{equation}
Without any loss of generality, $\delta$ may be chosen real by absorbing its 
phase into $\nu_\mu$ and $\nu_\tau$ and $\delta_0$ set equal to zero by 
redefining $m_0$ and the other $\delta$'s.  As a result,
\begin{eqnarray}
&& \sin^2 2 \theta_{atm} \simeq 1, ~~~ \Delta m^2_{atm} \simeq 4 \delta m_0^2, 
~~~ U_{e3} \simeq {i Im \delta'' \over \sqrt 2 \delta}, \\ 
&& \Delta m^2_{sol} \simeq 4 \sqrt {(\tilde \delta')^2 + 2 (Re \delta'')^2} 
m_0^2, ~~~ \tan \theta_{sol} \simeq {\sqrt 2 Re \delta'' \over \sqrt 
{(\tilde \delta')^2 + 2 (Re \delta'')^2} - \tilde \delta'},
\end{eqnarray}
where $\tilde \delta' = \delta' + (Im \delta'')^2/2\delta < 0$.

Thus this model explains $\theta_{23} \simeq \pi/4$ and predicts three 
nearly degenerate neutrino masses with neutrinoless double beta decay 
given by $|m_0|$.  Since $\delta$ is a radiative correction, it cannot be 
too large.  Given that $\Delta m^2_{atm}$ is known to be of order $10^{-3}$ 
eV$^2$, $m_0$ cannot be much smaller than about 0.3 eV.  Remarkably, this 
is also the upper limit on neutrino mass from the large-scale structure of 
the Universe \cite{univ} and possibly the value of $|m_0|$ as measured in 
neutrinoless double beta decay \cite{beta2}.

In the Standard Model, there are no flavor-changing leptonic interactions, 
thus $\delta = \delta'' = 0$ and Eq.~(47) does not lead to neutrino 
oscillations at all.  However, if there is some new physics which allows 
all the $\delta$'s to be nonzero, then Eq.~(47) can be realistic.  A recent 
detailed example \cite{hrsvv} is available in the context of supersymmetry 
with arbitrary soft supersymmetry breaking terms.

\subsection{Arbitrary Neutrino Masses}

In addition to $\xi_1$ transforming as \underline {1} under $A_4$, consider 
$\xi_2$, $\xi_3$, and $\xi_{4,5,6}$ transforming as \underline {1}$'$, 
\underline {1}$''$, and \underline {3} as well \cite{ma04}.  In that case, 
${\cal M}_\nu$ in the original basis is given by
\begin{equation}
{\cal M}_\nu = \pmatrix {a+b+c & 0 & 0 \cr 0 & a + \omega b + \omega^2 c & 
d \cr 0 & d & a + \omega^2 b + \omega c},
\end{equation}
where $a$ comes from $\langle \xi_1^0 \rangle$, $b$ from $\langle \xi_2^0 
\rangle$, $c$ from $\langle \xi_3^0 \rangle$, and $d$ from $\langle \xi_4^0 
\rangle$, assuming that $\langle \xi_5^0 \rangle = \langle \xi_6^0 \rangle 
= 0$.

In the basis where the charged-lepton mass matrix is diagonal, the neutrino 
mass matrix becomes
\begin{equation}
{\cal M}_\nu^{(e,\mu,\tau)} = U_L^\dagger {\cal M}_\nu U_L^* = \pmatrix 
{a + (2d/3) & b - (d/3) & c - (d/3) \cr b- (d/3) & c + (2d/3) & a - (d/3) 
\cr c - (d/3) & a - (d/3) & b + (2d/3)}.
\end{equation}
This matrix has one obvious eigenstate, i.e. $\nu_2 = (\nu_e + \nu_\mu + 
\nu_\tau)/\sqrt 3$ with the eigenvalue $m_2 = a+b+c$.  Let
\begin{equation}
U = \pmatrix {\sqrt{2/3} & 1/\sqrt 3 & 0 \cr -1/\sqrt 6 & 1/\sqrt 3 & 
-1/\sqrt 2 \cr -1/\sqrt 6 & 1/\sqrt 3 & 1/\sqrt 2},
\end{equation}
then in the basis defined by this transformation, i.e.
\begin{eqnarray}
\nu_1 &=& \sqrt {2 \over 3} \nu_e - {1 \over \sqrt 6} (\nu_\mu + \nu_\tau), \\ 
\nu_2 &=& {1 \over \sqrt 3} (\nu_e + \nu_\mu + \nu_\tau), \\ 
\nu_3 &=& {1 \over \sqrt 2} (-\nu_\mu + \nu_\tau),
\end{eqnarray}
the neutrino mass matrix of Eq.~(51) rotates to
\begin{equation}
{\cal M}_\nu^{(1,2,3)} = U^\dagger {\cal M}_\nu^{(e,\mu,\tau)} U^* = 
\pmatrix {m_1 & 0 & m_4 \cr 0 & m_2 & 0 \cr m_4 & 0 & m_3},
\end{equation}
where
\begin{equation}
\pmatrix {m_1 \cr m_2 \cr m_3 \cr m_4} = \pmatrix {1 & -1/2 & -1/2 & 1 \cr 
1 & 1 & 1 & 0 \cr -1 & 1/2 & 1/2 & 1 \cr 0 & -\sqrt 3/2 & \sqrt 3/2 & 0} 
\pmatrix {a \cr b \cr c \cr d}.
\end{equation}
In the limit $m_4=0$, Eq.~(56) is diagonal and $U$ becomes the neutrino 
mixing matrix of Eq.~(1) with the prediction $\tan^2 \theta_{12} = 1/2$, 
as well as $\sin^2 2 \theta_{23} = 1$ and $\theta_{13} = 0$.  This is of 
course a well-known $ansatz$ \cite{hps}, but has only just been derived 
from the symmetry of a complete theory, without arbitrary assumptions 
regarding its charged-lepton sector, in Ref.~[10].

Note that $m_{1,2,3}$ in the above are all arbitrary.  In other words, the 
mixing angles are determined without regard to the masses, just as in 
the quark sector.  There is however an important difference.  Whereas all 
quark mixing angles are zero, the lepton mixing angles are not.  Further 
small corrections from physics beyond the Standard Model, such as 
supersymmetry \cite{hrsvv,bdm99}, are of course necessary to modify these 
predictions to coincide with present data.

Experimentally, $|U_{e3}|$ is known to be small, i.e. $m_4 = (\sqrt 3/2)(c-b)$ 
may be considered small compared to $|b+c|$.  Now $\Delta m^2_{atm} >> 
\Delta m^2_{sol}$ implies that either $d \simeq (3/2)(b+c)$ or $d \simeq -2a 
-(b+c)/2$.  If $d \simeq (3/2)(b+c)$, then
\begin{equation}
m_{1,2} \simeq a + b + c, ~~~ m_3 \simeq -a + 2(b+c).
\end{equation}
If $d \simeq -2a -(b+c)/2$, then
\begin{equation}
m_{1,2} \simeq a + b + c, ~~~ m_3 \simeq -3a.
\end{equation}
Either one will allow a normal hierarchy or an inverted hierarchy or nearly 
degenerate masses.

If $m_4 \neq 0$, $\nu_1$ mixes with $\nu_3$, but $\nu_2$ remains the same. 
Let the new mass eigenstates be
\begin{equation}
\nu'_1 = \nu_1 \cos \theta + \nu_3 e^{i\delta} \sin \theta, ~~~ 
\nu'_3 = -\nu_1 e^{-i\delta} \sin \theta + \nu_3 \cos \theta,
\end{equation}
then the new mixing matrix $U$ has elements
\begin{eqnarray}
&& U_{e1} = \sqrt {2 \over 3} \cos \theta, ~~~ U_{e2} = {1 \over \sqrt 3}, ~~~ 
U_{e3} = - \sqrt {2 \over 3} e^{i\delta} \sin \theta, \\
&& U_{\mu 3} = - {1 \over \sqrt 2} \cos \theta + {1 \over \sqrt 6} e^{i\delta} 
\sin \theta = - {1 \over \sqrt 2} \sqrt {1 - {3 \over 2} |U_{e3}|^2} - 
{1 \over 2} U_{e3}.
\end{eqnarray}
Therefore, the experimental constraint \cite{react}
\begin{equation}
|U_{e3}| < 0.16
\end{equation}
implies
\begin{equation}
0.61 < |U_{\mu3}| < 0.77,
\end{equation}
or, using $\sin^2 2 \theta_{atm} = 4 |U_{\mu3}|^2 (1-|U_{\mu3}|^2)$,
\begin{equation}
0.94 < \sin^2 2 \theta_{atm} < 1.
\end{equation}
Similarly, using $\tan^2 \theta_{sol} = |U_{e2}|^2/|U_{e1}|^2$,
\begin{equation}
0.5 < \tan^2 \theta_{sol} < 0.52
\end{equation}
is obtained.  Whereas Eq.~(65) is well satisfied by the current data, 
Eq.~(66) is at the high end of the 2$\sigma$-allowed range centered 
at $\tan^2 \theta_{sol} \simeq 0.4$ \cite{sol}.

If future experimental measurements persist in getting a value of $\tan^2 
\theta_{sol}$ outside the range predicted by Eq.~(66), one possible 
explanation within the context of this model is through radiative 
corrections.  Just as Eq.~(39) is radiatively corrected to become Eq.~(47), 
Eq.~(51) may also get corrected so that $\nu_1$ mixes with $\nu_2$ in 
Eq.~(56).  For example, if $b,c,d < a$, then combining Eqs.~(47) and (51) 
with $b=c$,
\begin{equation}
\tan 2 \theta_{sol} \simeq -2 \sqrt 2 \left[ {b-(d/3)+\delta''a \over 
b-(d/3)-2\delta'a} \right],
\end{equation}
where
\begin{equation}
\delta'' = \delta_{e\mu} + \delta_{e\tau}, ~~~ \delta' = \delta_{ee} - 
{1 \over 2} (\delta_{\mu\mu} + \delta_{\tau\tau}) - \delta_{\mu\tau},
\end{equation}
and $\tan^2 \theta_{sol}\simeq 0.4$ is obtained if $[b-(d/3)+\delta''a]/
[b-(d/3)-2\delta'a] \simeq 0.75$.  Note that this may occur even if 
$\delta_{\alpha \beta}=0$ for $\alpha \neq \beta$, i.e. in the absence of 
flavor-changing radiative corrections, 
in contrast to the requirement of Refs.~[8] and [9].

\section{Conclusion}

The non-Abelian discrete symmetry $S_3$ is ideal for explaining maximal 
mixing in the $\mu-\tau$ sector with a normal hierarchy of neutrino masses. 
In a specific application \cite{cfm04}, it also explains why $U_{e3}$ is 
small but nonzero.

The non-Abelian discrete symmetry $A_4$ is a natural 
candidate for describing 3 families of quarks and leptons.  Whereas 
Dirac fermion masses come from the decomposition
\begin{equation}
\underline {3} \times (\underline {1} + \underline {1}' + \underline {1}'') 
= \underline {3},
\end{equation}
Majorana neutrino masses come from the decomposition
\begin{equation}
\underline {3} \times \underline {3} = \underline {1} + \underline {1}' + 
\underline {1}'' + \underline {3}.
\end{equation}
The mismatch between the quark mass matrices is then naturally given by 
$V_{CKM} = 1$, while that between the charged-lepton and neutrino mass 
matrices is definitely not the unit matrix, but rather $U_{MNS(P)}$ of 
Eq.~(1) in a certain symmetry limit, thus predicting a relationship among 
$\theta_{23}$, $\theta_{12}$, and $\theta_{13}$.  
Specifically, if $\theta_{13} = 0$, then $\sin^2 \theta_{23} = 1$ and 
$\tan^2 \theta_{12} = 0.5$, independent of the values of the 3 neutrino 
masses.  [Note that all 6 quarks and all 3 charged leptons have names, but 
the 3 neutrinos do not, as yet.]  To obtain small nonzero quark mixing angles 
as well as deviations from the lepton mixing angles constrained by this 
model, new physics beyond the Standard Model is expected, such as 
supersymmetry at the TeV scale.

\section*{Aknowledgment}

I thank the Institute for Particle Physics Phenomenology, Durham for 
hospitality.  This work was supported in part by the U. S. Department of 
Energy under Grant No. DE-FG03-94ER40837.

\bibliographystyle{unsrt}

\end{document}